\documentclass[aps,prl,twocolumn,showpacs,amsmath,floatfix,citeautoscript]{revtex4}
\usepackage{graphicx}
\usepackage{dcolumn}

\begin{document}
%\topmargin-0.8cm

% \draft

\title{
Interface enhancement of ferroelectricity in CaTiO$_3$/BaTiO$_3$ superlattices
}

\author{Xifan Wu$^{1}$}
\author{Karin M. Rabe$^{2}$, and David Vanderbilt$^{2}$ }
\affiliation{$^{1}$Department of Physics, Temple Materials
Institute, and Institute for Computational Molecular Science,
Temple University, Philadelphia, PA 19122, USA \\
$^2$Department of Physics and Astronomy, Rutgers University,
Piscataway, NJ 08854-8019, USA}

\date{\today}

\date{\today}

\begin{abstract}

We carry out first-principles calculations for CaTiO$_3$/BaTiO$_3$
superlattices with epitaxial strain corresponding to growth on a SrTiO$_3$
substrate, and consider octahedral rotations as well as
ferroelectric distortions.  The calculations are done as a
function of electric displacement field, and both a macroscopic
and a local electrostatic analysis are
carried out.  We find that strong octahedral rotations occur for
TiO$_6$ octahedra sandwiched between CaO layers on both sides, but
are strongly suppressed if either neighboring layer is a BaO
layer.  Due to the resulting enhancement of the ferroelectric instability
in the BaO-neighboring octahedra, we find that overall the ferroelectric
instability of the superlattice is enhanced by the interface.
Thus, short-period superlattices in this system have a larger ferroelectric
polarization than longer-period ones of the same average composition,
contrary to the expected trend.

%
% This first-principles modeling approach
% provides a powerful tool for the theoretical prediction and design
% of complex ferroelectric superlattice structures.
%
\end{abstract}


\maketitle

%% \narrowtext

\marginparwidth 2.7in

\marginparsep 0.5in

\def\dvm#1{\marginpar{\small DV: #1}}
\def\kr#1{\marginpar{\small KR: #1}}
\def\xwm#1{\marginpar{\small XW: #1}}
\def\scr{\scriptsize}

% \dvm{We may want to iterate on the title.}

Perovskite oxide superlattices are currently the subject of intense
scientific interest \cite{Sai_PRL, Jeff_APL_2003, Lee_Nature,
Ghosez_Nature}.
%
% \kr{We need to pay some attention to the choice of these and
% following references.  Are there any review articles?}
%
Through variation of the choice of constituent materials and their
order of deposition, as well as the epitaxial strain state
via selection of the substrate, a wide variety of functional
properties can be achieved.  While the properties can be understood
partly as the result of the behaviors of the constituent
materials under the electrical and mechanical boundary conditions
characteristic of the superlattice \cite{Jeff_APL_2003}, in
many cases the atomic and electronic features of the interfaces
have been shown to play a critical role \cite{Xifan_PRL_2008}. As
a result, materials design via
interface engineering is becoming an essential strategy in guiding
experimental exploration of this enormous class of materials,
and there is clearly a pressing need for improved atomic-scale
understanding of the role played by the interfaces in determining
the functional properties of perovskite superlattices.

% \dvm{\scr Since you are not using bibtex, you will have to remember to
% reorder the references in the order of citation when we are done.}
%\xwm{The references have been reorganized.}

First-principles methods have been extremely useful in providing
both qualitative and quantitative characterization of interface
effects in superlattices \cite{Xifan_PRL_2008, Xifan_PRL_2006}.
In most perovskite superlattices combining ferroelectric (FE)
and paraelectric (PE)
layers \cite{Xifan_PRL_2008,Jeff_APL_2003,
Serge_PRB_2006, Karen_PRB_2005, JunHee_JAP_2009, Lee_Nature},
%
% \kr{Should we include experimental references as well?}
%
the FE instability and spontaneous polarization are
found to be suppressed with increasing interface density for
a given overall composition. In an analysis based on
layer polarizations defined in terms of localized Wannier orbitals
\cite{Xifan_PRL_2006} and using electric displacement $D$ as
the fundamental field variable, the properties of interfaces
were found to be determined mainly by the adjacent atomic layers
\cite{Xifan_PRL_2008,Max_all}.
%
%\dvm{I added a citation to Max's Nature Materials paper and condensed
%the citations of all three of Max's papers into one bibitem.  Also
%added a couple of additional references to it later on in the paper.}
%
In this local analysis,
the suppression of ferroelectricity by interfaces was associated
with the formation of an interface dipole pointing antiparallel
to the local polarization.

An additional feature of many perovskite oxides is a tendency for
rotations and tilts to occur in the corner-sharing network
of oxygen octahedra. For example, the crystal structure of bulk
CaTiO$_3$ (CTO) is generated by large oxygen octahedral rotations;
its nonpolar character is thus understood as the result of the
competition between FE and antiferrodistortive (AFD) orders, which
suppresses the FE instability
in the AFD state \cite{Zhong_Ferroelectrics}. In constrast,
BaTiO$_3$ (BTO) is highly resistant to oxygen octahedral rotations and
exhibits a robust FE state at room temperature
\cite{Zhong_PRL_1994}.  Previous first-principles studies of the
CTO/BTO superlattice system have neglected the oxygen octahedral
rotations \cite{Xifan_PRL_2008, Serge_PRB_2006, Jo_PRL_2010}, yielding
an artificially strengthened ferroelectricity and an enhanced
ground-state polarization \cite{Xifan_PRL_2008, Serge_PRB_2006}
which increases as the interface density decreases. It is quite
surprising, then, that a
slight {\it increase} in polarization from the 2:2 to the 1:1 superlattice
is observed in a recent experiment \cite{Lee_APL_2009}.

In this paper we show that an increased density of
interfaces can in fact enhance the
polarization of a superlattice, illuminating the experimental observation
\cite{Jo_PRL_2010}.
The key to this unusual
behavior is the role played by octahedral rotations, in
addition to FE distortions, in determining the energetics
and polarization of the interfaces \cite{Ghosez_Nature}.
Working with superlattices comprised of CTO and BTO layers
as our model system, we use first-principles calculations to
demonstrate this novel effect and to clarify the reasons for its
occurrence.
We find that the octahedral rotations are large for TiO$_6$ octahedra
sandwiched between CaO layers, but are strongly suppressed if either
neighboring layer is a BaO layer.  The latter case results in
a local enhancement of the FE instability and of the
resulting polarization.  As a result, short-period superlattices,
in which the density of interfaces is higher, have a
larger spontaneous polarization than longer-period ones of the
same average composition.

The details of the calculations are as follows.  We carry out
strained-bulk calculations on all distinct period-four-compatible
superlattices built from
CaTiO$_3$ (C) and BaTiO$_3$ (B) cell layers, namely
BTO, 3B1C, 2B2C, 1B1C, 1B3C, and CTO,
stacked in the [001] direction.  The
in-plane lattice constant is fixed at $a_0$=$7.275$\,bohr, our computed
equilibrium lattice constant for bulk cubic SrTiO$_3$, corresponding to
epitaxial growth on SrTiO$_3$. We use a 40-atom
tetragonal supercell with lattice vectors of length $\sqrt{2}a_0$ in
the [110] and [1$\bar 1$0] directions
and $c \approx 4a_0$ along the [001] direction.
This allows us to consider both the FE distortion
along [001] and an antiferrodistortive pattern of octahedral rotations
about the [001] axis, resulting in $P4bm$ space-group symmetry.
We neglect the tilting of oxygen octahedra (rotations about an
in-plane axis); we believe this is justified because oxygen tilting
requires a coherent pattern of tilts that would propagate into the
BTO unit cells, where octahedral rotations are unfavorable.
%absent or that their effects should be weak.
%
The structural relaxations and electron minimizations were done at
fixed electric displacement fields
%in the range of $-$0.32 to 0.32\,C/m$^2$
within the framework of density-functional theory in the
local-density approximation \cite{LDA_PW92} using the
LAUTREC code package \cite{Max_all}, which
implements plane-wave calculations in the projector augmented-wave
framework \cite{PAW}.
We used a plane-wave cutoff energy of 80\,Ry and a 4$\times$4$\times$1
Monkhorst-Pack $k$ mesh.
%For local investigation of the interfaces, we carried out out layer
%polarization (LP) analysis as in
%Refs.~\onlinecite{Xifan_PRL_2006,Xifan_PRL_2008}.
%
\begin{table}
\caption{Computed tetragonal distortion ($c/4a_0$) and spontaneous
polarization, in C/m$^2$, of the CTO/BTO supercells, with (P$_s$)
and without (P$_{s,0}$) octahedral rotations.}
\begin{ruledtabular}
\begin{tabular}{lcccccc}
& BTO & 3B1C & 1B1C & 2B2C & 1B3C & CTO \\
\hline
$P_s$ & 0.388 & 0.333 & 0.184 & 0.161 & 0.0 & 0.0 \\
$c/4a$ & 1.0595 & 1.0359& 1.0114 & 1.0110 & 0.9939 & 0.9798 \\
$P_{s,0}$ & 0.388 & 0.346 & 0.266 & 0.374 & 0.447 & 0.590 \\

\end{tabular}
\end{ruledtabular}
\label{table:Polarization}
\end{table}

In Table I we present the spontaneous polarization $P_s$ and $c/a$ ratio
for each of the six superlattices considered. As expected, $P_s$
increases with increasing BTO fraction. Comparing the two superlattices
2B2C and 1B1C, which have the same overall composition, we find
that the latter, with higher interface density, has a larger
$P_s$, contrary to the expectations based on
FE-PE superlattices previously considered in the literature
\cite{Jeff_APL_2003, Xifan_PRL_2008, Serge_PRB_2006}, but consistent
with the experimental result reported in Ref.~\cite{Lee_APL_2009}.
This enhancement in 1B1C, however, is not due to improper 
ferroelectricity as was the case for PbTiO$_3$/SrTiO$_3$ described
in Ref.~\cite{Ghosez_Nature}. In both 1B1C and 2B2C, the rotations
suppress, rather than enhance, the polarization, as can be seen
by comparing the polarization in the presence of rotations with
the polarization computed for rotations constrained to zero,
given in the last row of Table I. Though the polarization in the
absence of rotations is much greater for 2B2C, reflecting the
expected trend in which a higher interface density suppresses the
polarization, the magnitude of the suppression by rotations is
much greater for 2B2C than for 1B1C, due to the large rotation
angle in the CC layer. The net result is to reverse the trend,
with the polarization in 1B1C with rotations being slightly larger
than for 2B2C. Application of a
large electric field, as in Ref.~\cite{Jo_PRL_2010}, is expected
to favor states with higher polarization and smaller octahedral
distortion, reflecting the competition between these two
distortions; this could be investigated by application of the
present
approach but will not be further discussed here.

\begin{figure}
\includegraphics[width=2.4in]{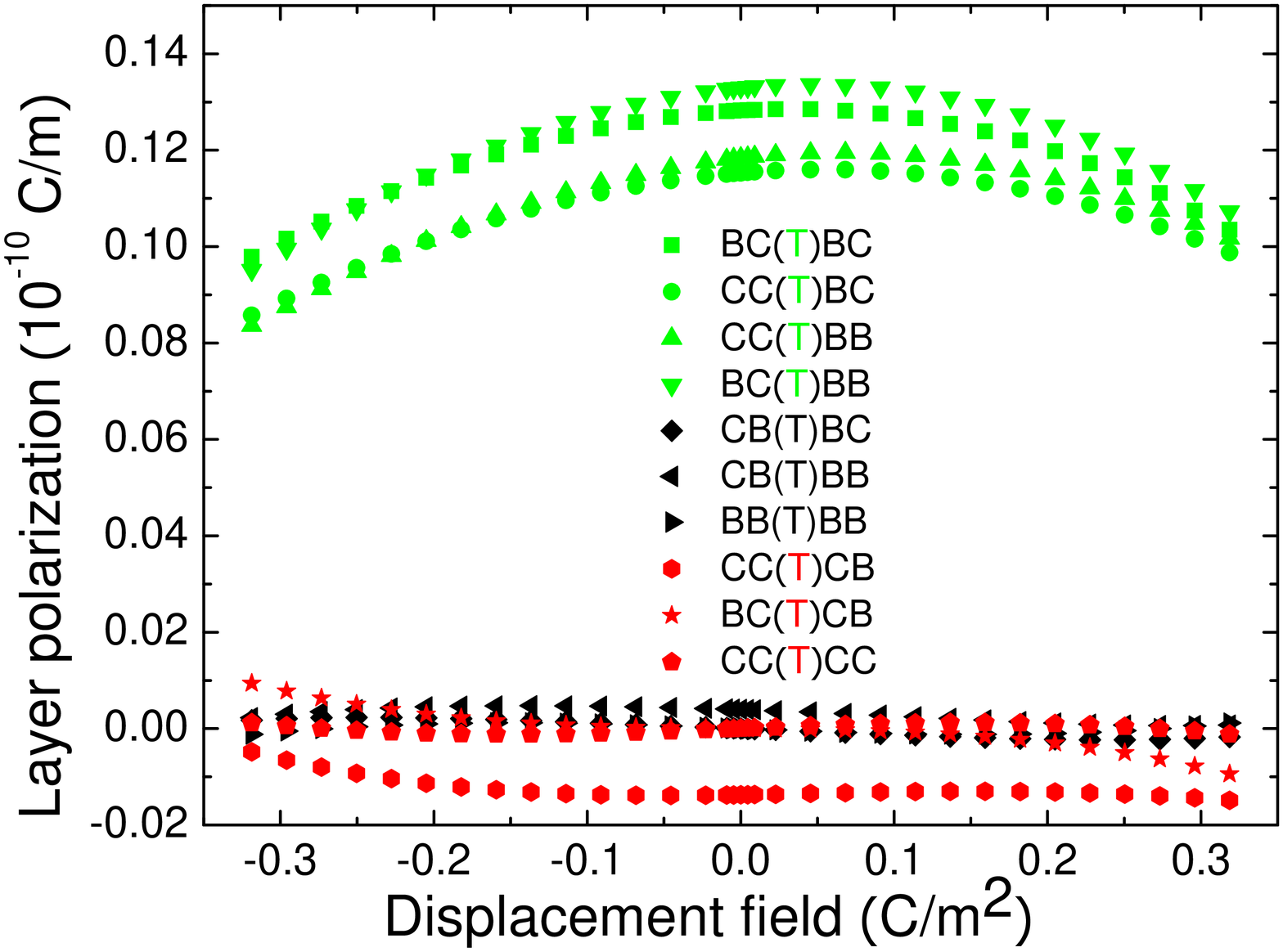}
\caption{\label{fig1} (Color online) Dependence of TiO$_2$ layer
polarizations (relative to the average of TiO$_2$ planes in bulk
BaTiO$_3$ and CaTiO$_3$) on chemical environment in CTO/BTO supercells.}
\end{figure}

To develop an atomic-scale understanding of the enhancement of
the polarization in 1B1C, we carry out a layer-by-layer analysis of
structural and dielectric properties of the
superlattices ~\cite{Xifan_PRL_2006,Xifan_PRL_2008}.
In previous work \cite{Xifan_PRL_2008},
we showed that the properties of each atomic layer
depend only on the chemical identity of neighboring layers and on global
parameters such as the in-plane lattice constant and the displacement
field $D$ (along $z$).   We expect this ``locality principle'' to continue
to apply when the AFD rotations are included.  In
Fig.~\ref{fig1}, we plot the layer polarizations of
TiO$_2$ layers as a function of $D$ for all the CTO/BTO supercells
considered.  For each $D$, we fully relax the structure, optimizing
with respect to rotations of the octahedra in each layer.
It can be seen that the layer polarizations are mainly
determined by the chemical identity of the first-neighbor layers,
just as was the case when octahedral rotations were constrained to
zero \cite{Xifan_PRL_2008}.

% \kr{add a comment on the dependence on D?}

%%%%%%%%%%%%%%%%%%%%%%%%%%%%%%%%%%%%%%%%%%%%%%%%%%%%%%%%%%%%
%%%%%%%%%%%%%%%%%%%%%%%%%%%%%%%%%%%%%%%%%%%%%%%%%%%%%%%%%%%%

We next consider the octahedral rotations and show that these, too,
are functions of the local chemical environment.
% in a similar way as for the layer polarizations.
In all of the CTO/BTO superlattices studied here, we found that
structures having rotations that change phase from layer to layer have lower
energy than those with in-phase rotations.  Henceforth we
characterize the structures by the absolute value $\theta_j$ of the
rotation angle for TiO$_2$ layer $j$, with the understanding that
the signs always alternate from layer to layer.

%%%%%%%%%%%%%%%%%%%%%%%%%%%%%%%%%%%%%%%%%%%%%%%%%%%%%%%%%%%%
\begin{figure}
\includegraphics[width=2.3in]{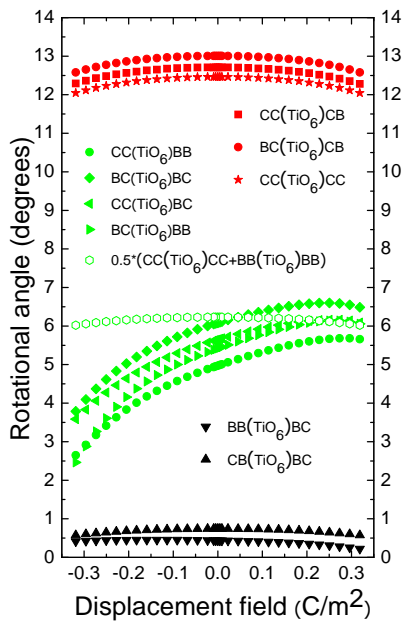}
\caption{\label{fig2} (Color online.) TiO$_6$ rotation angles as
a function of $D$ for TiO$_6$ octahedra sandwiched between
(a) two CaO layers, (b) one BaO and one CaO layer, and (c) two BaO
layers.}
\end{figure}
%%%%%%%%%%%%%%%%%%%%%%%%%%%%%%%%%%%%%%%%%%%%%%%%%%%%%%%%%%%%

The calculated dependence of the rotation angles on $D$ is presented in
Fig.~\ref{fig2}.  The results are grouped into three categories
depending on whether the two first-neighbor AO layers are both CaO,
one CaO and one BaO, or two BaO layers.  It is clear
that the octahedral rotations depend strongly on the identity of
the first-neighbor layers, while the influence of the second-neighbor
layers is much weaker.  A closer inspection shows that
a large octahedral rotation of $\sim$12-13$^\circ$ is observed
whenever the TiO$_6$ layer is sandwiched by two CaO layers,
while only a tiny rotation is found for the TiO$_6$ octahedra
sandwiched between BaO layers. This is consistent with the
properties of the parent materials, since rotations are
strongly favored in bulk CTO while being suppressed in bulk
BTO.  We find that the $D$-field dependence of the octahedral rotations
is generally weak in both cases.

What is more intriguing, however, is the behavior of the interfacial
layers of TiO$_6$ octahedra.  These experience a strongly broken mirror
symmetry due to the presence of BaO on one side and CaO on the other.
The rotation angles are intermediate in size,
$\sim$4-6$^\circ$, and are much more strongly dependent \cite{explan-Ddep}
on $D$.
(A similar effect due to second-neighbor asymmetry, barely visible
for some of the curves in Fig.~\ref{fig2}(a) and (c), has a very
much weaker amplitude.)
It is this reduction in the angle that allows a larger polarization
to develop in this layer, enhancing the overall polarization of the
system.

\begin{table}
\caption{Reduced inverse dielectric permittivities
$\epsilon_{\rm r}^{-1} = \epsilon_0/\epsilon$ computed at
$D$=0 for the bulk and supercell structures studied.}
\begin{ruledtabular}
\begin{tabular}{lcrr}
             & CaTiO$_3$ & Rotations & Rotations \\
Superlattice & fraction  & allowed\;   & forbidden \\
\hline
BTO	    & 0    & $-$0.0191~ & $-$0.0191~ \\
3B1C      & 0.25 & $-$0.0118~ & $-$0.0108~ \\
1B1C      & 0.50 & $-$0.0030~ & $-$0.0039~ \\
2B2C      & 0.50 & $-$0.0028~ & $-$0.0074~ \\
1B3C      & 0.75 & 0.0063~ & $-$0.0070~ \\
CTO       & 1    & 0.0157~ & $-$0.0151~ \\
\end{tabular}
\end{ruledtabular}
\label{table:invperm}
\end{table}

This behavior can also be analyzed by examining the tendency of a
system towards ferroelectric behavior as measured by the inverse
permittivity $\epsilon^{-1}= \partial E / \partial D$ evaluated at
$D$=0 \cite{Max_all}, where $E$ is the electric field.  For a PE
system $\epsilon^{-1}(D=0)$ is positive, and its approach to zero
signals the proximity to a FE instability; for a FE system it gives
a measure of the strength of the FE instability that is independent
of anharmonicity and thus complementary to the spontaneous polarization.

In the third column of Table \ref{table:invperm} we present
the inverse permittivities
$\epsilon^{-1}(D=0)$ for our bulk and superlattice structures
for the case that rotations are allowed.
As expected, the (strained) bulk BTO shows the most
FE behavior (strongly negative $\epsilon^{-1}$), while
the (strained) bulk CTO remains PE ($\epsilon^{-1}>0$).
The dielectric responses of the mixed superlattices are thus
intermediate between those of bulk BTO or CTO, as expected.
Consistent with the interface density dependence of the
polarization, we find that the FE instability of the 1B1C
superlattice is slightly greater than that of the 2B2C
superlattice, and both of these are greater than the average of
BTO and CTO, which corresponds to
$n$B$n$C in the limit $n\rightarrow\infty$ and interface density
equal to zero.

The emergence of this interface enhancement effect is related to the
coexistence of FE and AFD order in the system.  To demonstrate this,
the last column of Table \ref{table:invperm} presents the
inverse dielectric permittivities with octahedral rotations
constrained to zero.
Here, we find that bulk CTO has nearly the same degree of FE
instability as bulk BTO, while the presence of interfaces reduces
the FE instability substantially for the 3B1C, 2B2C,
and 1B3C superlattices, and even more for the 1B1C
superlattice, with a two-fold increase in interface density.
Comparing with the values when rotations are allowed, it is clear that
the suppression of the octahedral rotation greatly increases
the FE tendency in the supercells involving 50\% or more CTO.
%

% Leave blank space before "From" so mailer convert to ">From"
%
 From these comparisons, it follows that the competition
between AFD and FE orders is essential for the interface enhancement effect
in this system.  In fact, there is a local relationship between the
octahedral rotations in a layer and their contribution to the FE
instability. For the purposes of carrying out a local decomposition,
we focus on the inverse of the capacitance per basal area,
$C^{-1}=\partial V/\partial D$, at $D$=0. Here $V=LE$ is
the potential drop across the supercell, with $L$ being the
supercell height in the stacking direction and
$E$ the average electric field.
Apart from a small correction proportional to $\partial L/\partial D$,
which we have verified has little effect on the analysis, this is
equal to $L\epsilon^{-1}$, and thus can be used
as a measure of ferroelectric instability similar to
$\epsilon^{-1}$ \cite{Max_all}.
We decompose $C^{-1}$ locally as follows.
First, for each AO or
TiO$_2$ layer in the superlattice, we define an inverse layer capacitance
\begin{equation}
c_j^{-1}=\frac{1}{\epsilon_0}\left(h_j+D\frac{\partial h_j}{\partial D}
- \frac{\partial p_j}{\partial D}\right)
\end{equation}
where $h_j$ and $p_j$ are the layer height \cite{Layer_height}
and layer polarization \cite{Xifan_PRL_2006} associated with
each layer $j$.  We then assign an inverse cell capacitance to the cell
centered on TiO$_2$ layer $i$ as
\begin{equation}
C^{-1}_i= c^{-1}_{{\rm TiO_2},i} +
 \frac{1}{2}c^{-1}_{{\rm AO},i-1}
+ \frac{1}{2}c^{-1}_{{\rm AO},i+1}.
\end{equation}

\begin{table}
\caption{Inverse cell capacitance (Jm$^2$/C$^2$) at $D=0$ for each
TiO$_2$-centered unit cell in the 2B2C supercell,
for the case that rotations are allowed ($P4bm$ symmetry) or forbidden
($P4mm$ symmetry).}
\begin{ruledtabular}
\begin{tabular}{lccccc}
& $C^{-1}_{\rm B(T)B}$ & $C^{-1}_{\rm B(T)C}$ & $C^{-1}_{\rm C(T)C}$
    & $C^{-1}_{\rm C(T)B}$ & $C^{-1}_{\rm tot}$ \\
\hline
$P4bm$ & $-$0.416 & $-$0.361 & $\phantom{-}$0.663 & $-$0.361 & $-$0.475 \\
$P4mm$ & $-$0.007 & $-$0.152 & $-$0.985  & $-$0.152 & $-$1.296 \\
\end{tabular}
\end{ruledtabular}
\label{table:Inverse_capacitance}
\end{table}

Taking the 2B2C supercell as an example, we report the resulting inverse
cell capacitances at $D=0$ in Table \ref{table:Inverse_capacitance}.
The last column of the table is $C^{-1}_{\rm tot} =\sum_{i=1}^{4}C_i$.
For comparison, the results are given both for the case that rotations
are allowed and when they are forbidden.  As expected, the absence of
the oxygen rotations results in a much stronger FE instability,
as shown by the fact that $C^{-1}_{\rm tot}$ is much more negative
than when rotations are allowed.
The individual inverse cell capacitance show changes in Table I that
are not always easy to interpret, because the inverse capacitance of
a given TiO$_2$-centered unit cell does still depend fairly strongly
on the identity of neighboring cells.  For example,
$C^{-1}_{\rm B(T)B}$ is not independent of environment, despite the
fact that the rotations in that TiO$_2$ layer are quite small.
However, there is one huge change that stands out in Table I, namely,
that $C^{-1}_{\rm C(T)C}$, the inverse capacitance of the cell containing
the octahedra sandwiched between CaO layers, changes drastically
when rotations are allowed.
Here, the oxygen rotation angle is $\sim$13$^\circ$, similar
to what it is in bulk CTO, and this induces an enormous
suppression of the FE instability in that layer,
switching the local tendency from a ferroelectric to a strongly
PE one.

These observations help explain the result presented earlier
in the paper: that short-period superlattices can be more FE than 
long-period ones.  We can expect that
there are two competing influences: (i) the effect
that typically occurs in FE-PE perovskite oxide superlattices in which
the presence of an interface interrupts the FE order
and reduces the FE tendency; and (ii) an effect that
is driven by oxygen octahedral rotations and arises because of the severe
suppression of ferroelectricity that occurs only for TiO$_6$
octahedra sandwiched between Ca layers on both sides.  The density
of such layers in an $n$B$n$C superlattice is $(n-1)/2n$, or $1/2n$
less than in the average of bulk BTO and CTO.  This $1/2n$
reduction of a suppressing influence amounts to an enhancement
of the FE tendency in short-period superlattices.
Effects (i) and (ii) are of opposite sign but evidently have
similar magnitudes, with (ii) dominating. This accounts for the increase 
in polarization for the increased interface density of the 1:1 superlattice 
relative to the 2:2 superlattice, as observed in a recent
experimental investigation \cite{Lee_APL_2009}.

In conclusion, we have shown that oxygen octahedral rotations must
be included for any realistic modeling of CTO/BTO superlattices.  Using
first-principles calculations, we find that the octahedral rotations
in a given layer depend strongly on
the identity of the neighboring layers.  In particular, the
rotational angle in a TiO$_2$ layer sandwiched between two CaO layers
is found to be very large, while one between two BaO layers rotates
hardly at all, similar to the tendencies of the corresponding
bulk CTO and BTO parent materials.  The competition between
rotations and ferroelectricity then plays out locally, and
results in a novel behavior in which the ferroelectricity is found to
increase systematically with decreasing $n$ in $n$B$n$C superlattices,
in contrast to what has been found when only the FE
degrees of freedom are present.  These results have critical implications
for the design and application of short-period superlattice structures
having enhanced or novel properties.

\acknowledgments

We acknowledge M.~Stengel for suggesting the form of the inverse
layer capacitance used in the local dielectric analysis, and
thank H.N.~Lee for useful discussions.  The work was supported by
ONR grants N00014-05-1-0054 and N00014-09-1-0302.

%%%%%%%%%%%%%%%%%%%%%%%%%%%%%%%%%%%%%%%%%%%%%%%%%%%%%%%%%%%%


\begin{thebibliography}{0}
\bibitem{explan-Ddep}
The positive slope results from the fact that $D<0$ pushes
the interfacial TiO$_6$ octahedra into closer contact with the
BaO layer, which tends to suppress rotation.
\bibitem{Layer_height}
The height $h_j$ of layer $j$ is defined as
$(\bar{Z}_{j+1} - \bar{Z}_{j-1})/2$, where $\bar{Z}_j$
is the average [001] coordinate of all ions belonging to layer $j$.
\end{thebibliography}
\end{document}